\documentclass{czjphys}         
\usepackage{graphicx}

\begin{document}
\title{The Laser Calibration System of the ALICE Time Projection Chamber}  
%
\authori{G. Renault, B.S. Nielsen, J. Westergaard and J.J. Gaardh\o je for the ALICE Collaboration.}      \addressi{Niels Bohr Institute Blegdamsvej 17, DK-2100 Copenhagen \O}
\authorii{}     \addressii{}
\authoriii{}    \addressiii{}
\authoriv{}     \addressiv{}
\authorv{}      \addressv{}
\authorvi{}     \addressvi{}
%
\headauthor{G. Renault et al.} 
\headtitle{The Laser Calibration System \ldots} 
\lastevenhead{G. Renault et al.:  \ldots} 

\refnum{A}
\daterec{XXX}    
\issuenumber{0}  \year{2001}
\setcounter{page}{1}
\maketitle

\newcommand{\dNdy}{dN/dy}

\begin{abstract}
A Large Ion Collider Experiment (ALICE) is the only experiment
at the Large Hadron Collider (LHC) dedicated to the
study of heavy ion collisions. The Time Projection Chamber (TPC)
is the main tracking detector covering the pseudo rapidity range $|\eta|< 0.9$.
It is designed for a maximum multiplicity \dNdy = 8000.
The aim of the laser system is to simulate ionizing tracks at predifined
positions throughout the drift volume in order to monitor the TPC response to
a known source. In particular, the alignment of the read-out chambers
will be performed, and variations of the drift velocity due
to drift field imperfections can be measured and used as calibration data in
the physics data analysis. 
In this paper we present the design of
the pulsed UV laser and optical system,
together with the control and monitoring systems.
\end{abstract}

\begin{figure}[htb!]
\center
\includegraphics[width=\textwidth]{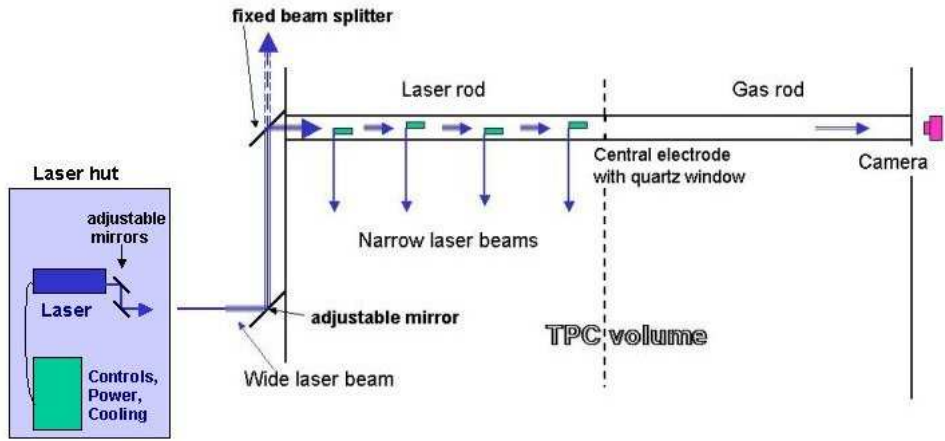}
\caption{The TPC laser principle. A 25 mm wide and 5 ns long
pulse hits an adjustable mirror and is guided inside the TPC.
The beam is splitted into 1 mm-wide rays in a rod
to simulate ionizing particle tracks.
}
\label{FtpcLaserPrinciple}
\end{figure}

The ALICE experiment will study heavy ion collisions at LHC.
The main tracking detector of the ALICE experiment is the TPC\cite{ALICETDRTPC2000}.
The aim of the laser system is to generate straight tracks
at known positions
in the drift volume of the TPC. 
These tracks will be generated by two-photon
ionization of the drift gas
by a pulsed UV laser beam with a wave length of 266 nm.
Other electrons are emitted
by photoelectric effect when the laser beam hits metallic surfaces
such as the central electrode, the aluminized mylar
strips of the electric field cage, or wires and pads of the readout system.
After readout and track reconstruction using the TPC detector,
distorsions related to ExB effects and mechanical
misalignment will be measured and corrected using these tracks.
The spatial and temporal variations of
the drift velocity due to the drift field will be measured within
a relative error of $10^{-4}$ and used as calibration data
in the physics analysis.

\begin{figure}[htb!]
\center
\includegraphics[]{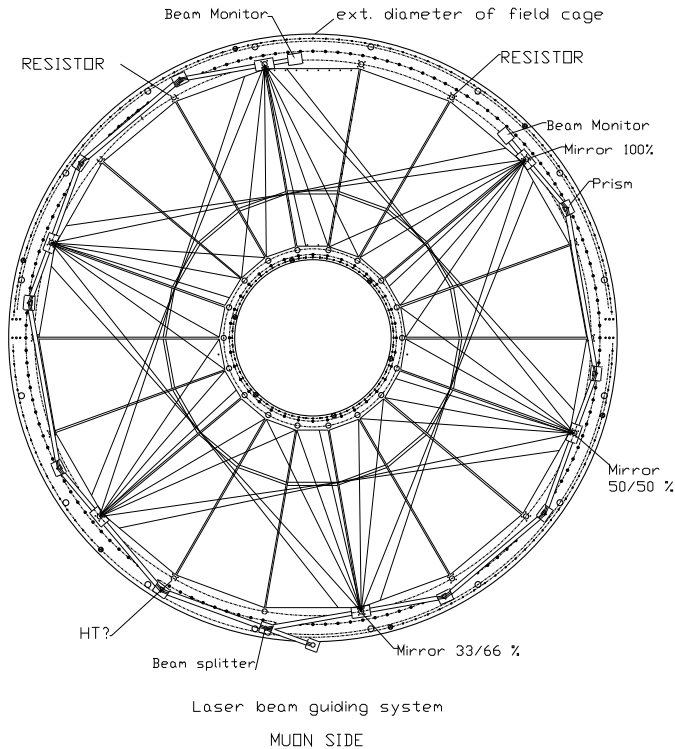}
\caption{336 laser tracks cross the TPC volume at predefined positions and directions.}
\label{Flasertracks}
\end{figure}

The calibration system
is composed of a static optical system and of adjustable parts.
The static optics is composed of beam splitters, mirrors and bending prisms
guiding the laser beam around the TPC field cage before it enters the TPC volume.
The guiding system ends with cameras that are able to take pictures
of the laser beam in order to monitor the position and the beam intensity.  
The adjustable part is mainly composed of remotely adjustable mirrors that will guide
the beam into the static optics system.
In order to generate multiple tracks in the TPC, the laser beam
goes through several steps. 
First,
a 25 mm diameter laser beam is divided into 6 secondary beams by
splitters before entering the TPC from one side (Fig. \ref{Flasertracks}).
Each beam enters the TPC in a rod (Fig. \ref{FtpcLaserPrinciple}),
passes through the central
electrode and, for monitoring purposes, is detected by a camera
located on the other side of the TPC.
In each rod, there are
4 fixed micro-mirror bundles.
Each of them extracts
seven 1 mm-diameter beams (20-40 $\mu$J/pulse)
from the secondary beam and reflects them as rays into the drift volume.
A second laser beam
generates similar rays in the second half of the TPC.
Thus 336 laser beams on the whole (Fig. \ref{Flasertracks})
will be created inside the TPC
volume \cite{LASERCALIB2002}.
\begin{figure}[htb!]
\center
\includegraphics[width=6cm,height=8cm]{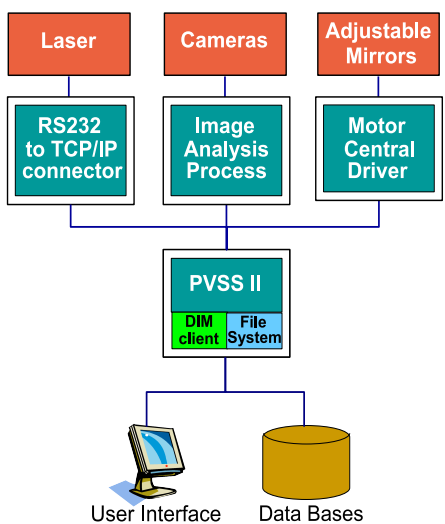}
\caption{Sketch of the Calibration Laser Control System.}
\label{LASERDCS}
\end{figure}

The actual laser is from Spectron Laser Systems Ltd,
model SL805-UPG.
The RS232 laser connections are converted into a TCP/IP connection
with an interface from Digi International\cite{DIGI} company in order
to be able to control the laser over long distances.    
The laser is controled by a C++ serial port
driver included in a DIM\cite{DIM} server under Windows XP.
This server communicates with a DIM client included
in the user interface.
The laser delivers 5 ns pulses at a repetition rate of 10Hz.
The energy
of a pulse is 100 mJ for a wave length of 266 nm
at the entrance of the end cap. A second laser will be
installed in the laser hut, at z $\approx -15$ m.
Both lasers can be used
in parallel, one for each end cap, or one for both
end caps, the other one being kept as a backup \cite{LASERCALIB2002,LASERDCS}.

Two adjustable mirrors are used to align the laser
beam in the guiding system on the TPC endplates.
They are controlled by New Focus hardware and
software for TCP/IP connections.

Cameras (up to 16) will monitor the laser
beams. At each end cap one camera is placed behind
the movable mirror. Two others can be placed at
the end of the two $180^{\circ}$ bends.
Moreover, 12
extra cameras are at the end of each rod.
All these cameras will take pictures of the laser beam
in order to monitor its intensity and its position. 

A synchronization board
will synchronize laser pulses and cameras with the LHC clock.
This board was developed for the Detector Control System
of the TPC and TRD detector in ALICE\cite{ALICETDRTPC2000,ALICETDRTRD2001}.

The Calibration Laser Control System (Fig. \ref{LASERDCS})
is set up to provide an easy
and safe structure to use the laser from
the control room.
It will be integrated into the Detector
Control System of ALICE.
Images of
the laser spot recorded by cameras
can be processed in order to move mirrors
to align laser beams inside rods.
All sub-systems (the laser, cameras
and adjustable mirrors) will be controlled
by a common Supervisory Control
And Data Acquisition called PVSS
(Prozessvisualisierungs-und Steuerungs-System) 
inside the Join
Controls Project framework\cite{PVSS}, giving a
trustable user interface.

As a conclusion, all optical elements will be set up and aligned
in the TPC from november 2005 to february 2006. This includes
micro-mirror bundles in rods, mirrors and prisms in support wheels,
adjustable mirrors and cameras.
The laser will be operational in spring 2006, ready
for generating multiple tracks for
TPC tests with cosmic rays in summer 2006.



\begin{thebibliography}{99}

\bibitem{ALICETDRTPC2000}
ALICE Collaboration:
ALICE Technical Design Report of the Time Projection Chamber,
CERN/LHCC 2000-001 (2000). 

\bibitem{LASERCALIB2002}
B.S. Nielsen, J. Westergaard, J.J. Gaardh\o je, A. Lebedev and ALICE TPC Collaboration:
Design Note on the ALICE TPC laser calibration system,
ALICE-INT-2002-022 (2002).

\bibitem{DIGI}
http://www.digi.com/

\bibitem{DIM}
http://dim.web.cern.ch/dim/

\bibitem{LASERDCS}
D. Beck: 
DCS User Requirement an Architectural Design
Specifications for the Calibration Laser Sub-System for the ALICE TPC.

\bibitem{ALICETDRTRD2001}
ALICE Collaboration:
ALICE Technical Design Report of the Transition Radiation Detector,
CERN/LHCC 2001-021 (2001). 

\bibitem{PVSS}
http://itcobe.web.cern.ch/itcobe/Services/Pvss/



\end{thebibliography}
\end {document}